\documentclass[twocolumn,showpacs,preprintnumbers,amsmath,amssymb]{revtex4}

\usepackage{graphicx}
\usepackage{dcolumn}
\usepackage{bm}
\usepackage[usenames]{color}

\begin{document}

\preprint{}

\title{Cyclotron resonance and mass enhancement by electron correlation in KFe$_2$As$_2$}

\author{Motoi Kimata$^{1,2\dagger}$, Taichi Terashima$^{1,2}$, Nobuyuki Kurita$^{1,2}$, Hidetaka Satsukawa$^1$,
Atsushi Harada$^1$, \\Kouta Kodama$^{1*}$, Kanji Takehana$^1$, Yasutaka Imanaka$^1$, Tadashi Takamasu$^1$, Kunihiro Kihou$^{2,3}$, Chul-Ho Lee$^{2,3}$, Hijiri Kito$^{2,3}$, 
Hiroshi Eisaki$^{2,3}$, Akira Iyo$^{2,3}$, Hideto Fukazawa$^{2,4}$, Yoh Kohori$^{2,4}$, Hisatomo Harima$^{2,5}$, and Shinya Uji$^{1,2}$}
 \altaffiliation[Also at ]{the Graduate School of Pure and Applied Sciences, University of Tsukuba. \\
 $^{\dagger}$Present address: Institute for Solid State Physics, University of Tokyo, Kashiwa, Chiba 277-8581, Japan. }
 
\affiliation{$^1$National Institute for Materials Science (NIMS), Ibaraki 305-0003, Japan}%
\affiliation{$^2$JST, Transformative Research-Project on Iron Pnictides (TRIP), Tokyo 102-0075, Japan. }%
\affiliation{$^3$National Institute of Advanced Industrial Science and Technology (AIST), Ibaraki 305-8568, Japan. }
\affiliation{$^4$Department of Physics, Chiba University, Chiba 263-8522, Japan. }
\affiliation{$^5$Department of Physics, Graduate School of Science, Kobe University, Hyogo 657-8501, Japan. }



\begin{abstract}

Cyclotron resonance (CR) measurements for the Fe-based superconductor KFe$_2$As$_2$ are performed. One signal for CR is observed, and is attributed to the two-dimensional $\alpha$ Fermi surface at the $\Gamma$ point. We found a large discrepancy in the effective masses of CR [(3.4$\pm$0.05)$m_e$ ($m_e$ is the free electron mass)] and de-Haas van Alphen (dHvA) results, a direct evidence of mass enhancement due to electronic correlation. A comparison of the CR and dHvA results shows that both intra- and interband electronic correlations contribute to the mass enhancement in KFe$_2$As$_2$.

\end{abstract}

\maketitle


Recently discovered Fe-based superconductors \cite{kamihara}  have attracted much attention because of their high transition temperatures ($T_c$). The maximum $T_c$ exceeds 54$-$56 K \cite{kito,ren,wang}. 
For Fe-based superconductors, substantial evidence points to the occurrence of unconventional superconductivity (extended $s$-wave and nodal gap structures) both experimentally and theoretically \cite{ishida,johnstona}. 
In unconventional superconductivity, it is believed that a strong electronic correlation exists, which renormalizes the mass of the conduction electrons, and that it is closely related to the superconducting mechanism. 
Because effective mass is an excellent measure of electronic correlation for a conductor, the effective masses as well as the Fermi surface (FS) structures provide valuable information on the superconducting mechanism. 
Thus far, the FS structures and effective masses ($m^*$) for Fe-based superconductors have been reported in various experiments, such as those involving de-Haas van Alphen (dHvA) oscillations, optical spectroscopy, angle-resolved photoemission spectroscopy (ARPES) \cite{ishida,johnstona}, and angle-dependent magnetoresistance oscillations (AMROs) measurements \cite{kimata1}.
Recently, systematic dHvA measurements for BaFe${_2}$[As$_{\rm (2-x)}$P$_{\rm x}$] revealed critical-like increase of the mass enhancement factor toward the antiferromagnetic (AF) phase boundary \cite{shishido}, which is another important aspect needed for understanding the superconducting mechanism.

KFe$_2$As$_2$ is an end member of the [Ba$_{\rm (1-x)}$K$_{\rm x}$]Fe$_2$As$_2$ family. This material has a relatively low $T_c$ ($\approx$3 K) and shows no structural or magnetic transitions \cite{rotter, chen}. 
Therefore, we can study original large FSs, whose structure is closely related to the superconducting mechanism. 
This is in a sharp contrast to the other end member BaFe$_2$As$_2$, in which a magnetic transition associated with structural change takes place at around 140 K \cite{rotter2}, and reconstructed small FSs are observed at low temperatures \cite{terashima3}. 
The specific heat, thermal conductivity, microwave penetration depth, and small-angle neutron scattering measurements for KFe$_2$As$_2$ indicate the presence of a nodal superconducting gap structure \cite{fukazawa, dong, hashimoto2,kawano-furukawa}. 
However, the detailed gap structure is still under debate: $d$-wave symmetry with vertical line nodes or extended s-wave symmetry with horizontal nodes is proposed \cite{hashimoto2, kawano-furukawa}. Because the optimally doped material of this family (x $\approx0.4$) is considered as a fully gapped superconductor, the study of the FS structure and effective mass of KFe$_2$As$_2$ has a significant meaning for the understanding of the superconducting mechanism in [Ba$_{\rm (1-x)}$K$_{\rm x}$]Fe$_2$As$_2$ family. According to the ARPES measurements \cite{sato, tyoshida}, there are large FSs ($\alpha$, $\zeta$, and $\beta$) centered at the $\Gamma$ point and small $\epsilon$ FSs near the corners of the first Brillouin zone (FBZ), although there are discrepancies in their sizes. 
In the dHvA measurement \cite{terashima}, three different FSs ($\alpha$, $\zeta$, and $\epsilon$) are found, whose effective masses are  6.0 and 6.5$m_e$ ($\alpha$), 8.5 and 18$m_e$ ($\zeta$), and 6.0 and 7.2$m_e$ ($\epsilon$) for the minimal and maximal cross sections of the quasi-two-dimensional (Q2D) FS ($m_e$ is the free electron mass), respectively. 
The corresponding mass enhancement factors (=$m^*_{\rm dHvA}/m_{\rm band}$, where $m_{\rm band}$ is the calculated band mass) are reported to be: 4.3 and 2.7 ($\alpha$), 3.9 and 6.9 ($\zeta$), and 20 and 24 ($\epsilon$), respectively.
The AMRO measurements reveal two series of oscillations: strong oscillations associated with $\alpha$ and a weak one associated with $\zeta$ \cite{kimata1}. 
For the $\alpha$ FS, the averaged mass enhancement factor from the dHvA measurements (=3.5) is reasonably consistent with the recently reported ARPES measurements result, where $m^*_{\rm ARPES}/m_{\rm band}\approx3$ \cite{tyoshida}.  
These values are rather large compared to the mass enhancement factors observed in another Fe-based superconductor LaFePO ($m^*_{\rm dHvA}/m_{\rm band}\approx2$) \cite{coldea}. 
Such a large mass enhancement is likely due to the strong electron-electron (e-e) interaction in KFe$_2$As$_2$. The electron-phonon (e-p) coupling in iron-pnictides is generally thought to be weak as suggested by theoretical estimations of the coupling constants for LaOFeAs and BaFe$_2$As$_2$ \cite{yildirim,boeri, boeri2}.
However, there still exist some inconsistencies in the FS structures for different experiments, which renders the arguments explaining mass enhancement less reliable. 

Here, we report the cyclotron resonance (CR) experiments for KFe$_2$As$_2$. 
Single crystals of KFe$_2$As$_2$ are obtained from the same batch as that used for our previous dHvA and AMRO measurements. 
Although CR measurements are a powerful method to directly obtain the effective mass of a conductor, 
they cannot determine the size of the cross sectional area and the position of the FS in the $k$-space, leading to some ambiguity in assignment of the CR signal for multi FS systems. In this study, we have compared the CR results with our previous AMRO and dHvA measurements, which allows the unambiguous assignment of the CR signal.
A comparison of the effective mass determined from CR measurements ($m^*_{\rm CR}$) with $m^*_{\rm dHvA}$ further enables us to determine the contribution of e-e interaction to mass enhancement \cite{kohn, kanki, hill1, ardavan, myoshida}.  We obtain the e-e coupling constants corresponding to the intra- and interband e-e interactions separately, which are closely related to the superconducting mechanism.


\begin{figure}
\includegraphics[width=7cm]{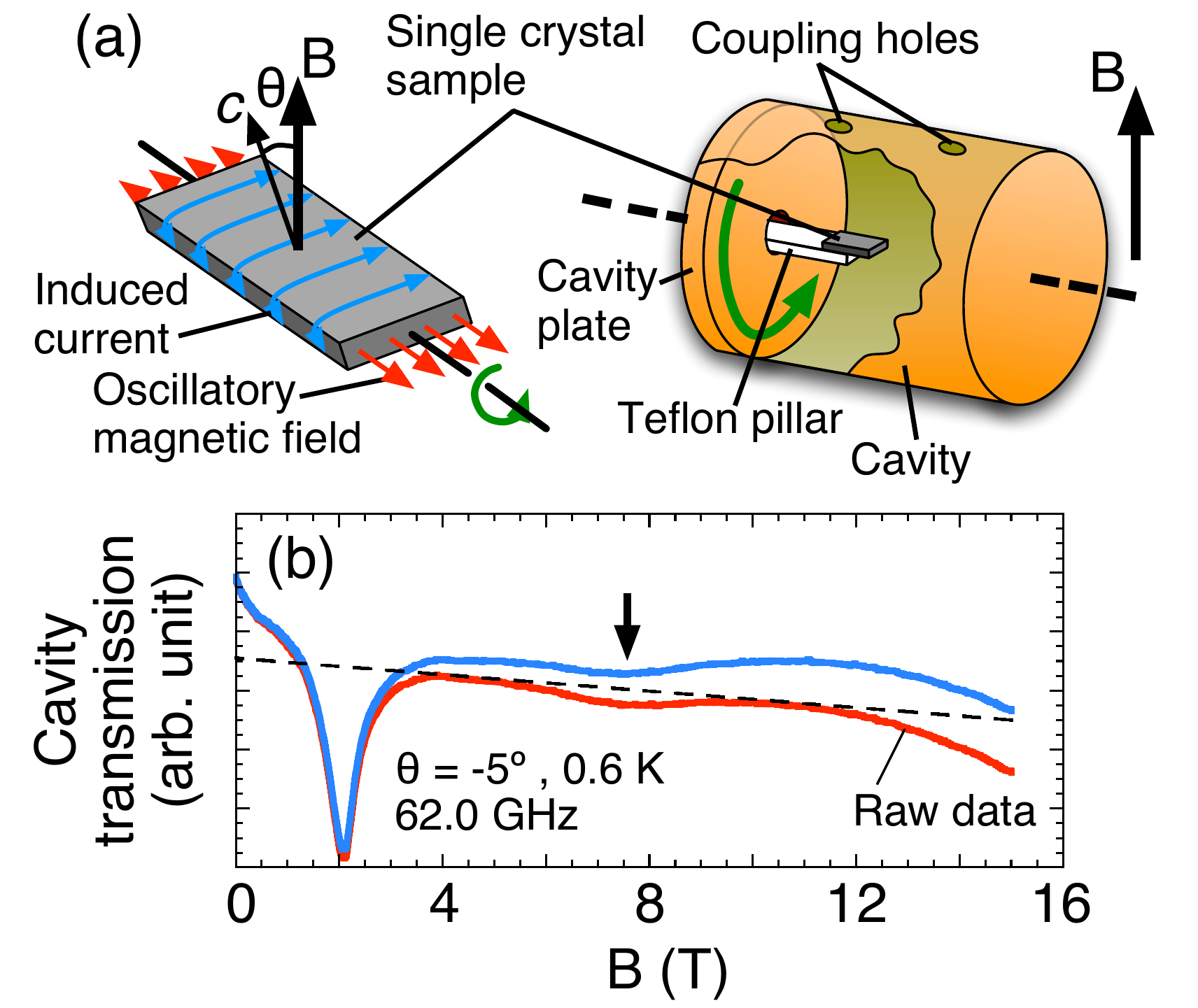}
\caption{\label{f1}(color online). (a) Schematic of the sample and cavity configuration. The induced ac current is coupled with the cyclotron motion of electrons. 
(b) Magnetic field dependence of the cavity transmission for $\nu$ = 62.0 GHz and $T$ = 0.6 K. The lower (red) and upper (blue) curves show the raw data and the background-subtracted data, respectively (see text for details). The sharp absorption at around 2 T is attributed to the ESR of some impurities or contaminations, and the broad absorption (indicated by the arrow) corresponds to CR. }
\end{figure}

Single crystals of KFe$_2$As$_2$ are synthesized by a flux method \cite{kihou}. 
The residual resistivity ratios (=$\rho$(300K)/$\rho$(5K)) of the crystals amount to approximately 600. The plate-like crystals with typical dimensions of 1 $\times$ 2 $\times$ $<$(0.1) mm$^3$ are used for the measurements. The interlayer direction ($c$-axis) is perpendicular to the basal plane of the crystal [Fig. 1(a)]. The CR experiments are performed in the range from 55 GHz to 82 GHz by a cavity perturbation technique \cite{myoshida, mvna}. 
The sample, which is mounted on a small Teflon pillar with a small amount of grease, can be rotated in a $^3$He cryostat \cite{kimata2, takahashi}, and a dc magnetic field of up to 15 T is applied by a superconducting solenoid [Fig. 1(a)].
The TE$_{01n}$ ($n$ = 1$-$3) modes are used, consequently, the ac magnetic field is applied parallel to the long axis of the sample at any field angle [Fig. 1(a)].
The resistivity of the present sample at 5 K along the $ab$-plane ($\approx1$ $\mu\Omega$cm) and the $c$-axis ($\approx20$ $\mu\Omega$cm) gives the microwave penetration depths for $\nu$ = 60 GHz as $\delta_{ab}\approx$ 0.2 $\mu$m and $\delta_{c}\approx$ 0.9 $\mu$m, respectively. They are much smaller than the sample dimensions: the measurements are done in the skin-depth regime.


The lower (red) curve in Fig. 1(b) shows the magnetic field dependence of the cavity transmission for $\nu$ = 62.0 GHz (TE$_{012}$ mode)  at 0.6 K. 
The upper critical field at low temperatures ($T$$<$0.1 K) for this field direction is approximately 1.6 T \cite{terashima2}.
We subtract a smooth background [straight dashed line in Fig.1 (b) ], likely arising from the dc magnetoresistance of the sample and the cavity surface, from the raw data, and obtain the upper (blue) curve shown in Fig. 1(b). 
Similar background subtractions are employed in all the following analyses. 
The sharp resonance (dip) at around 2 T is almost independent of both the temperature and magnetic field angle, but has a strong sample dependence. Hence, this signal is attributed to the electron spin resonance (ESR) of some impurities or contaminations.
Actually, recently performed low-temperature specific heat measurements reveal the presence of unknown magnetic impurities in the crystals \cite{kim}.  
The broad absorption observed at around 7.5 T [arrow in Fig. 1(b)] is attributed to the intrinsic CR signal of the sample because of the systematic dependence of this absorption on the temperature, frequency, and field angle, as shown later. 
All the transmission data in the following figures are vertically shifted for clarity. 
Figure 2 shows the field dependence of the transmission at various temperatures. 
A series of broad CR signals at around 7.5 T are clearly observed below 3 K, where $k_BT<\sim$$h\nu$, and they become more prominent as the temperature is lowered. 
The resonance field is almost independent of the temperature. 
The transmission drops rather abruptly above 12 T at low temperatures. 
The reason for this is not clear.

\begin{figure}
\includegraphics[width=6cm]{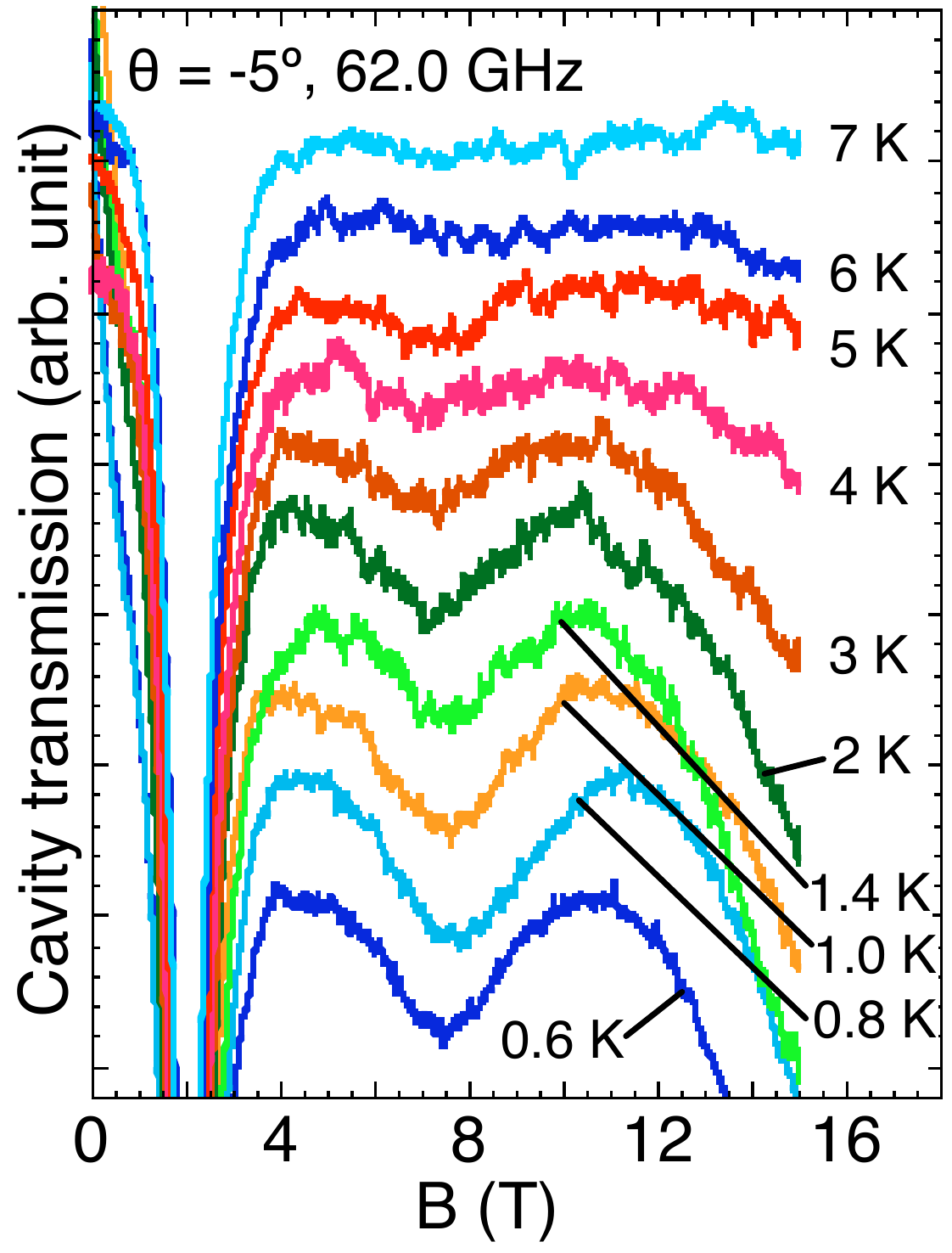}
\caption{\label{f1}(color online).  Field dependence of the cavity transmission at various temperatures. }
\end{figure} 

Figure 3(a) shows the field dependence of the transmission at various field angles $\theta$.
The resonance fields are determined by fitting the resonance curves to Eq. (8) in Ref. \cite{dresselhaus}. 
The resonance shifts to a higher field as $\theta$ increases, following a 1/cos$\theta$ dependence [Fig. 3(b)]. 
This clearly shows that the observed CR signals originate from the 2D FS of KFe$_2$As$_2$. 
The resonance also shifts to a higher field as the frequency $\nu$ increases [Fig. 4(a)]. 
The linear fit of the $\nu$ dependence gives the effective mass of (3.4$\pm0.05)m_e$ [Fig. 4(b)].

\begin{figure}
\includegraphics[width=5.5cm]{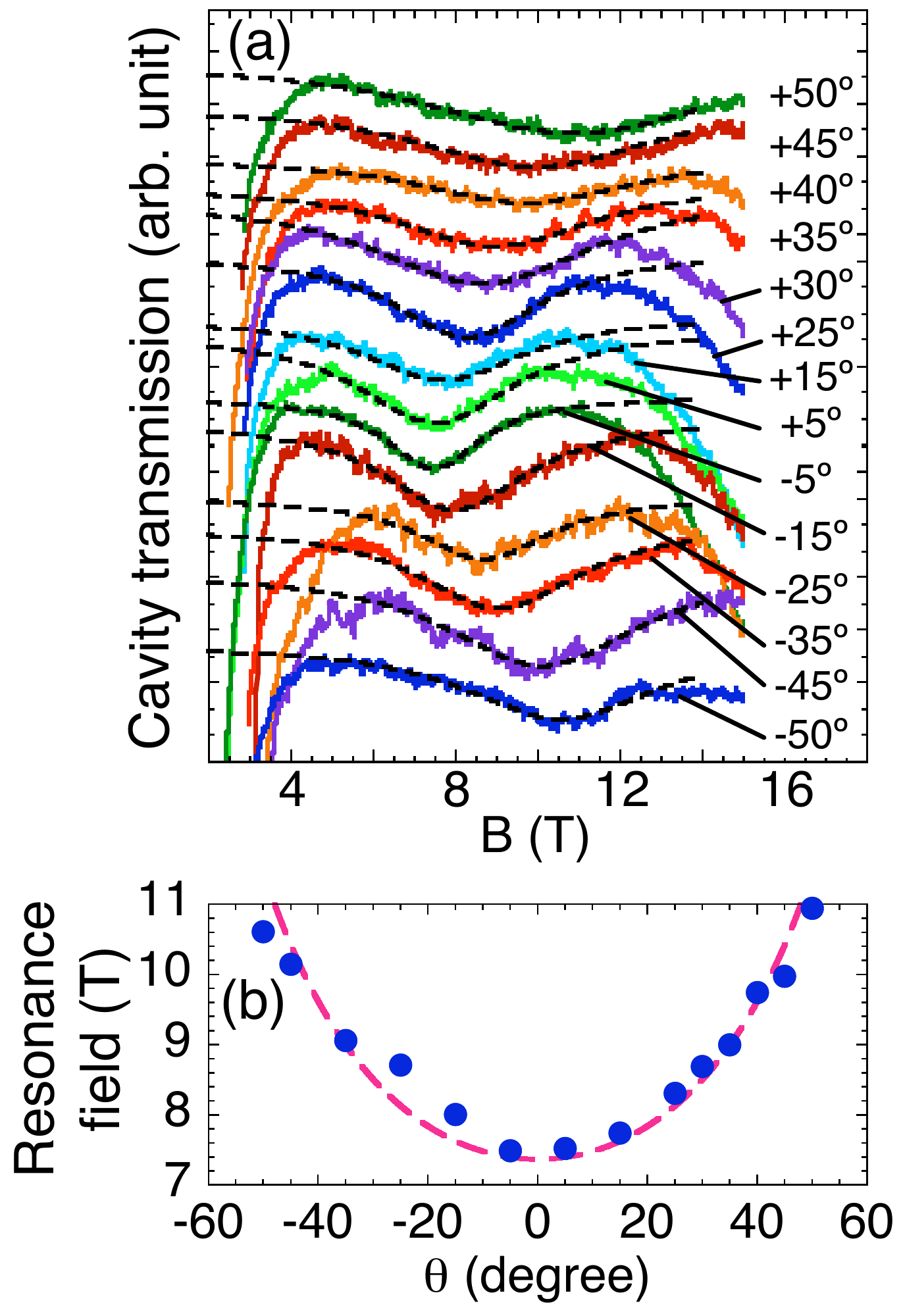}
\caption{\label{f1}(color online). 
(a) Field dependence of the transmission spectra at 0.6 K and $\nu$ = 62.0 GHz. The dashed curves indicate the fitting results obtained by Eq. (8) in Ref. \cite{dresselhaus}. (b) $\theta$ dependence of the resonance field. The dashed curve represents the $B_{\rm res}/{\rm cos}\theta$ with $B_{\rm res}$ = 7.4 T.}
\end{figure} 

\begin{figure}
\includegraphics[width=8cm]{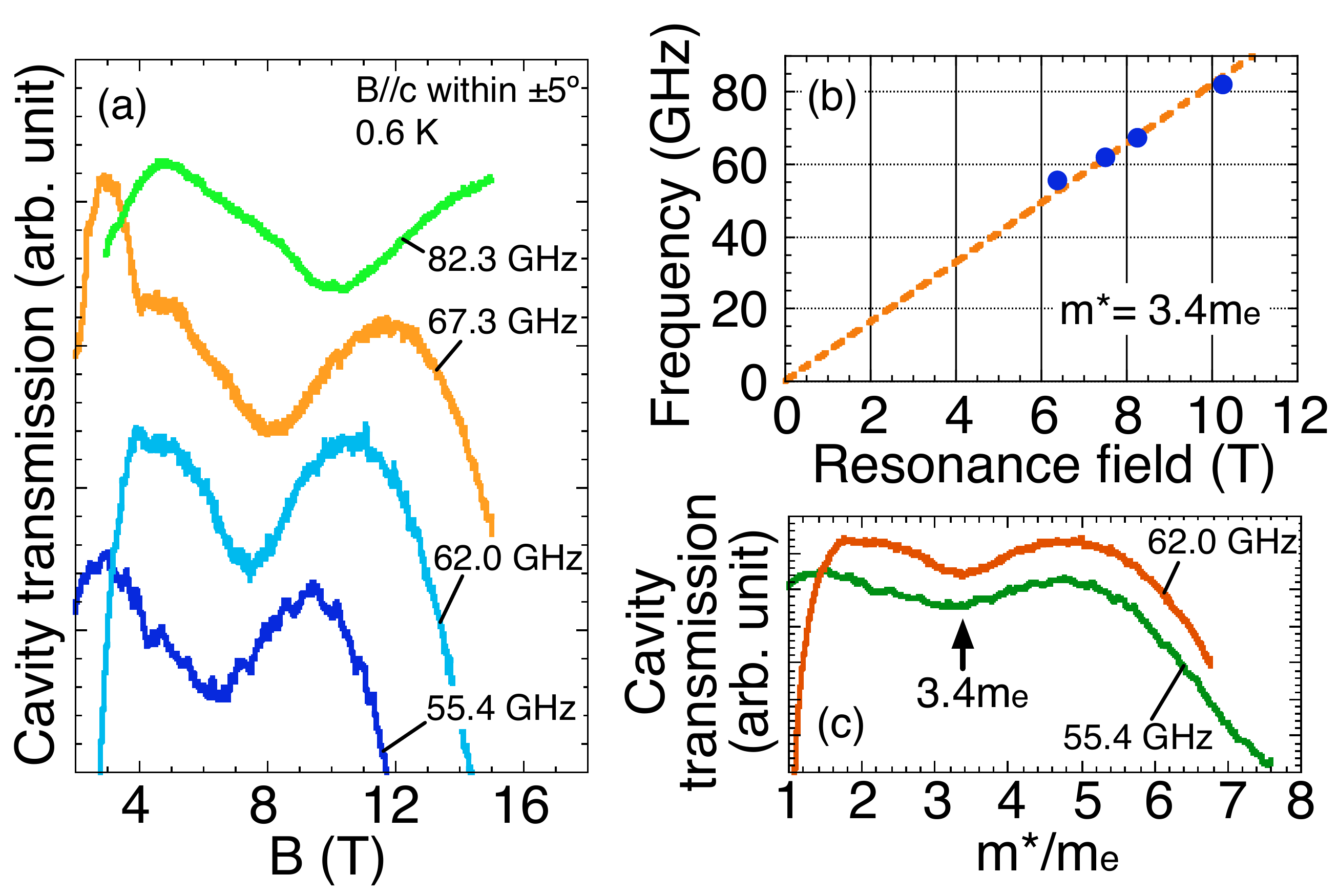}
\caption{\label{f1}(color online). 
(a) Transmission for several microwave frequencies at 0.6 K. (b) Frequency vs resonance field plot obtained from the data in (a). The dashed straight line gives the cyclotron effective mass of (3.4$\pm0.05$)$m_e$. (c) Transmission for $\nu$ = 55.4 and 62.0 GHz as a function of $m^*/m_e$.
Both spectra show only single resonance at 3.4$m_e$.}
\end{figure}

Although the multiple Q2D FSs have been observed in recent dHvA and AMRO measurements on KFe$_2$As$_2$, we have found only one series of resonances in the present CR measurement. 
Essentially the CR signal corresponds to the electrical conductivity under the high-frequency electric field. 
Therefore, the CR data are directly compared with the AMRO data, which are related to the electrical conductivity in the dc limit. 
In AMRO measurements, the $\alpha$ oscillation is well resolved and its amplitude is much stronger than that of $\zeta$. 
Therefore, the CR signal is likely ascribable to the $\alpha$ FS. 
Indeed, the scattering times obtained from the CR line width, $\tau_{\rm CR}\approx$ 0.6$-$1.4 $\times$ $10^{-11}$ s ($\omega_c\tau\approx4$), are consistent with those of the $\alpha$ FS obtained during the dHvA measurements, $\tau_{\rm dHvA}\approx$ 0.6$-$0.7 $\times$ $10^{-11}$ s. 
On the other hand, the $\zeta$ FS has a relatively three-dimensional shape (large energy dispersion along the $c$-axis) and the masses corresponding to the maximal and minimal cross sections are 8.5 and 18$m_e$, respectively \cite{terashima}. 
Therefore, the CR signal from the $\zeta$ FS is possibly broadened by the mass distribution and is consequently difficult to observe. 

The harmonic resonances of the fundamental CR are occasionally observed in some Q2D systems \cite{hill1, ardavan, oshima, kimata3}. 
The resonance fields of the harmonic CRs are given by $B_0/n$, where $B_0$ and $n$ are the resonance field of the fundamental CR and integers ($n=2, 3, 4, ...$), respectively. 
The origin of high-order harmonic CRs is related to the periodic motion of the electrons on the Q2D FS with higher order corrugations \cite{mckenzie, hill2, blundell}. 
The transmission spectra for $\nu=55.4$ GHz and 62.0 GHz at 0.6 K [Fig. 4(c)], where the horizontal axis is normalized by $m^{*}$/$m_e$, clearly show that only the fundamental CR is present but that no high harmonic CRs, e.g., no signals exist at 1.7 or 6.8$m_e$.

Here, we discuss the difference in the mass renormalization effects between the dHvA oscillation and CR measurements; $m^*_{\rm dHvA}=6.3m_e$ on an average, and $m^*_{\rm CR}=3.4m_e$ for the $\alpha$ FS.
It is well known that $m^*_{\rm dHvA}$ is renormalized by both e-p and e-e interactions. 
On the other hand, according to Kohn's theorem \cite{kohn}, a long-wavelength radiation can couple only to the center-of-mass motion of electrons, which is not affected by e-e interactions.
In this situation, corresponding to the CR measurements in single-band systems, $m^*_{\rm CR}$ is renormalized only by the e-p interaction (not by the e-e interaction). 
However, this is not the case for CR measurements in multi-band systems. 
In multi-band systems such as KFe$_2$As$_2$, $m^*_{\rm CR}$ can be renormalized by the interband (not intraband) e-e interaction \cite{myoshida, kublbeck, takada}.
Based on the above discussion, we assume the effective mass in KFe$_2$As$_2$ as $m^*_{\rm dHvA}=(1+\lambda_{\rm ep})(1+\lambda_{\rm ee}^{\rm intra}+\lambda_{\rm ee}^{\rm inter})m_{\rm band}$ and $m^*_{\rm CR}=(1+\lambda_{\rm ep})(1+\lambda_{\rm ee}^{\rm inter})m_{\rm band}$  \cite{prange}, where $\lambda_{\rm ep}$, $\lambda_{\rm ee}^{\rm intra}$, and $\lambda_{\rm ee}^{\rm inter}$ are the e-p, intraband e-e, and interband e-e coupling constants, respectively. 
For the $\alpha$ FS, we obtain $\lambda_{\rm ee}^{\rm intra}/(1+\lambda_{\rm ee}^{\rm inter})=0.9$ using $m^*_{\rm dHvA}/m_{\rm e}=6.3$ on an average and $m^*_{\rm CR}/m_{\rm e}$ = 3.4. 
If we use the theoretical value for the related materials, i.e., $\lambda_{\rm ep}\approx0.2$ \cite{boeri2, yildirim, boeri}, we get $\lambda_{\rm ee}^{\rm intra}=1.4$ and $\lambda_{\rm ee}^{\rm inter}=0.5$. This leads us to conclude that both the intra- and the interband e-e interactions enhance the effective mass of KFe$_2$As$_2$ although the $\lambda_{\rm ep}$ and $m_{\rm band}$ values remain rather ambiguous.

In Fe-based superconductors, a possible superconducting mechanism is the AF fluctuation between different FS pockets, as discussed in the literatures \cite{kuroki, mazin}. 
In KFe$_2$As$_2$, the main Q2D FSs are located at the $\Gamma$ point \cite{sato, tyoshida}, and a nodal gap structure has been proposed \cite{fukazawa, dong, hashimoto2,kawano-furukawa}. 
These might suggest that the superconductivity in KFe$_2$As$_2$ is mainly mediated by the intraband AF fluctuation \cite{hashimoto2}. However, the present CR study shows the existence of appreciable interband e-e interaction in KFe$_2$As$_2$ despite the absence of nesting between electron and hole FSs.  This may be consistent with a recent inelastic neutron scattering measurement, which suggests significant interband scattering between the bands around the $\Gamma$ and $X$ points \cite{lee}. At present, it is still controversial which AF fluctuation is the dominant superconducting mechanism 
for KFe$_2$As$_2$.

In the compensated Fe based superconductors, BaFe$_2$[As$_{\rm(2-x)}$P$_{\rm x}$], and LaFePO, significant discrepancies between the dHvA results and the band calculations are reported \cite{shishido, coldea}. The origin has been discussed in terms of the energy band shifts due to strong interband e-e interaction between the hole and electron FSs \cite{ortenzi}. In KFe$_2$As$_2$, the strong interband and/or intraband e-e correlation evidenced by the CR measurements might also be the origin of the discrepancies between the dHvA results and the band calculation \cite{terashima}. 

In summary, a single CR signal is observed in KFe$_2$As$_2$, which is ascribed to the $\alpha$ FS centered at the $\Gamma$ point. The obtained $m^*_{\rm CR}$ of (3.4$\pm$0.05)$m_e$ is significantly smaller than the $m^*_{\rm dHvA}$ of the $\alpha$ FS, i.e., 6.3$m_e$ on an average. This directly indicates the presence of the strong mass enhancement on account of the e-e interaction. A comparison between the dHvA and CR analyses results also suggests that the effective mass of KFe$_2$As$_2$ is enhanced by both the intra- and the interband e-e interactions.

This work was supported by grant-in-aid for Scientific Research on Innovative Areas, (No. 20102005) from JSPS, Japan. K.K. thanks NIMS for providing the junior research assistantship.


\bibliography{apssamp}

\end{document}